\documentclass[aps,prl,preprint, titlepage,showpacs]{revtex4}

\usepackage{graphicx}

\usepackage{dcolumn}

\usepackage{bm}

\begin{document}
\title{Shear flows and shear viscosity in a two-dimensional Yukawa system (dusty plasma)}
\author{V.~Nosenko\footnotemark[1] and J.~Goree}
\footnotetext[1]{Electronic mail vladimir-nosenko@uiowa.edu}
\affiliation {Department of Physics and Astronomy,
The University of Iowa, Iowa City Iowa 52242}
\date{\today}

\begin{abstract}
The shear viscosity of a two-dimensional liquid-state dusty plasma
was measured experimentally. A monolayer of highly charged polymer
microspheres, with a Yukawa interaction, was suspended in a plasma
sheath. Two counter-propagating Ar$^+$ laser beams pushed the
particles, causing shear-induced melting of the monolayer and a
shear flow in a planar Couette configuration. By fitting the
particle velocity profiles in the shear flow to a Navier-Stokes
model, the kinematic viscosity was calculated; it was of order
$1~\rm{mm}^2\rm{s}^{-1}$, depending on the monolayer's parameters
and shear stress applied.
\end{abstract} 
 \narrowtext

\maketitle

A broad range of charged particle systems can be modeled as Yukawa
systems that can have a liquid state; these include colloids,
certain dense astrophysical plasmas, and strongly coupled dusty
plasmas. A dusty plasma is a suspension of highly charged
micron-size particles in a plasma. When these particles are
confined, their mutual repulsion causes them to self-organize in a
structure called a plasma crystal, which can be in a crystalline
or liquid state.

Shear viscosity is a dynamic property of fluids required to
describe shear flows and damping of waves. Recently, molecular
dynamics (MD) simulations \cite
{Murillo01,Salin02&03,Hamaguchi02} and theory \cite
{Murillo03} have been used to predict the viscosity of a liquid
Yukawa system. While viscosity is commonly measured in colloids,
there are no reports of comparable measurements in dusty plasmas.
A dusty plasma is much softer than a colloid, and between its
particles it has only a rarefied medium consisting of free
electrons and ions, as well as neutral gas, which applies a
frictional drag to moving particles.

A plasma crystal is an analogue for molecular matter; the
particles represent molecules \cite {Morfill04}, but they have
the advantage of allowing direct imaging and thus a measurement of
their individual positions and velocities \cite {Melzer00}.
Confinement of particles is provided by natural electric fields in
the plasma. The particle suspension can be manipulated using
various forces. In Ref.~\cite {Konopka00rotation}, ion drag
forces were used to apply a shear stress that melted a
two-dimensional (2D) crystallized dusty plasma. In Ref.~\cite
{Morfill04}, a 3D dusty plasma in a liquid state flowed past an
obstacle where there was a shear. In those two experiments,
experimenters had no direct control over the forces and the
resulting shear. Here, we report an experiment using laser
radiation to apply a controlled and localized shear stress to a
monolayer lattice. The lattice melted, and we modeled the
resulting velocity profile to measure the nonequilibrium
viscosity. The geometry resembled a laminar planar Couette flow.

We used the apparatus of Ref.~\cite {Nosenko04}, with essentially
the same parameters, including an Ar pressure of $5$~mTorr.
A monolayer of microspheres was suspended in the plasma. The
particles had a diameter of $8.09\pm0.18$~$\mu$m \cite {Liu03}
and a mass $m=4.2 \times 10^{-13}$~kg. The particle suspension's
diameter was $50-60$~mm. The interparticle potential for particles
arranged in a single plane, like ours, was experimentally shown
\cite {Konopka00Yukawa} to be nearly Yukawa
$U(r)=Q(4\pi\epsilon_0r)^{-1}{\rm exp}(-r/\lambda_D)$, where $Q$
is the particle charge and $\lambda_D$ is the screening length.
(This should not be confused with quasi-2D arrangements of
particle chains, which might have a different potential.) The
particle suspension is characterized by $\kappa=a/\lambda_D$ and
${\it \Gamma}=Q^2/4\pi\epsilon_0akT$, where $T$ is the particle
kinetic temperature. For liquids, the characteristic length $a$ is
the 2D Wigner-Seitz radius \cite {Kalman04}; it is related to the
lattice constant $b$ for a perfect triangular lattice by
$a=(\sqrt{3}/2\pi)^{1/2}b$. To vary $a$, $\kappa$, and $Q$, while
keeping the plasma parameters constant, we used a different number
of particles in each of three experiments, Table~1. We used the
pulse technique of Ref.~\cite {Nosenko02} to measure $\kappa$ and
$Q$.

Initially, our suspension was an undisturbed triangular lattice,
Fig.~\ref {Fig. 1}(a). As is typical for a 2D monolayer with our
experimental conditions, it had a highly ordered state. The static
structure factor $S(\bf{k})$ has the distinctive peaks of a
hexagonal crystal, Fig.~\ref {Fig. 1}(b), and the pair correlation
function $g(r)$ has many peaks and a correlation length of $r_{\rm
corr}=(23-30)a$.

The particles were imaged through the top window by a video
camera. We digitized movies of $256$ frames at $30$ frames per
second. The $23.1\times17.3$~mm field of view included $370-770$
particles, Fig.~\ref {Fig. 1}(a). Coordinates $x,y$ and velocities
$u_x,u_y$ were then calculated \cite {Melzer00} for each particle
in each frame.

To apply a shear stress in a planar Couette configuration, in an
ordinary fluid one uses moving plates, but in our dusty plasma,
Fig.~\ref {Fig. 1}(c), we used counter-propagating laser beams, as
in Ref.~\cite {Lin I04}, but with more laser power. Particles are
pushed by the radiation pressure force $F$, which was verified
\cite {Liu03} to be proportional to an incident laser intensity.
We chose to use two opposing laser beams, rather than one, to
avoid introducing a net momentum to the particle suspension. An
Ar$^+$ laser beam, with a power that was varied up to $3.41$~W,
was split in two. At their foci at the particle location, the
laser beams had a diameter of $0.61$~mm at the $1/e^2$ level. Two
rapidly oscillating scanning mirrors rastered the beams into
vertical sheets, which made footprints on the lattice by striking
it at an angle of $6.7-8.8^\circ$ with respect to a horizontal
plane. The sheets extended beyond the edges of the suspension.

In discussing our results, we first note that steady-state
conditions in the velocity profiles developed within $3$~s after
turning on the laser. All the maps and profiles of particle
velocity and temperature presented here were computed well after
that time.

The particle velocity was always highest within the laser
footprints. The velocity diminished elsewhere, with a profile that
we will model.

Beginning with an undisturbed lattice, we applied increasing
levels of shear stress, and we observed that the particle
suspension passes through four stages elastic deformation, defect
generation while in a solid state, onset of plastic flow, and
fully developed shear flow. We present data for the latter two
stages. At the onset of plastic flow, Fig.~\ref {Fig. 1}(c), the
particles hopped between equilibrium lattice sites. Domain walls
developed, and they moved continuously. The crystalline order of
the lattice in the shearing region deteriorated, broadening the
peaks in the static structure factor $S(\bf{k})$ in Fig.~\ref
{Fig. 1}(d). At still higher levels of shear stress, the lattice
fully melted everywhere, and a shear flow developed, Fig.~\ref
{Fig. 1}(e). The particle motion was highly irregular on a small
scale compared to the interparticle spacing, but on a larger
scale, it was like a laminar flow in a fluid. The liquid-like
order of the particle suspension is clearly indicated by the
diffusiveness of the structure factor $S(\bf{k})$ in Fig.~\ref
{Fig. 1}(f). Particles were confined so that after flowing out of
the field of view they circulated around the suspension's
perimeter and re-entered the field of view, and the suspension did
not buckle in the vertical direction. Within the field of view,
more than $95\%$ of the time-averaged flow velocity was directed
in the $x$-direction, with less than $5\%$ of the flow velocity
diverted in the $y$-direction.

We used a method of binning the velocity measurements $u_x$ and
$u_y$ according to a particle's $y$ position to compute spatial
profiles $v_x(y)$ for the flow velocity, and the fluctuations
parameterized as the kinetic temperature $T_x(y)$ and $T_y(y)$.
Within each bin, we computed moments of the velocity, and then
time-averaged over 256 consecutive frames, beginning after a
steady state was attained. This yielded $T_x(y)$ and $T_y(y)$ from
the second moments of $u_x$ and $u_y$, respectively, as shown in
Figs.~\ref {Fig. 1}(c) and \ref {Fig. 1}(e); and $v_x(y)$ from the
first moment, as shown in Fig.~\ref {Fig. 2}(c). We used the
region between the laser sheets to calculate a spatially and
temporally averaged $<T_y>$, which we used to compute ${\it
\Gamma}_y$. We use ${\it \Gamma}_y$ and not ${\it \Gamma}_x$ to
characterize our system, because there was no external energy
input in the $y$-direction. Separately, we also computed maps of
velocity, $\bar{u}_x(x,y)$, which were time-averaged to reduce
noise, Figs.~\ref {Fig. 2}(a) and \ref {Fig. 2}(b).

At the onset of plastic flow, the velocity maps $\bar{u}_x(x,y)$
reveal interesting $120^\circ$ zigzag features, Fig.~\ref {Fig.
2}(a). This angle corresponds to a triangular lattice. The
underlying movement of individual particles is best seen by
viewing the movie \cite {Movie_URL}.

In the fully melted condition, the velocity map in Fig.~\ref {Fig.
2}(b) has no significant variation in the $x$ direction. In this
shear flow, particles slip past their neighbors \cite {Movie_URL}.

The flow velocity profile $v_x(y)$ in Fig.~\ref {Fig. 2}(c) was
curved, unlike in a traditional planar Couette flow where $v_x(y)$
is linear with $y$. We attribute this curvature to the frictional
drag exerted on particles by the gas. A balance between this drag
in the $x$ direction and the viscous transport of particle
momentum $mu_x$ in the $y$ direction away from the laser
footprints accounts for the observed steady-state velocity
profile.

We modeled the velocity profile in the continuum approximation,
using the Navier-Stokes equation with a term for the gas drag
$\partial {\bf v}/\partial t+({\bf v}\nabla){\bf
v}=-\rho^{-1}\nabla p+(\eta/\rho) \nabla ^2 {\bf
v}+[\zeta/\rho+\eta/(3\rho)]\nabla(\nabla\cdot{\bf v})-\nu_d {\bf
v}$. Here, the parameters for the continuum representing our
particle suspension are ${\bf v}$, $p$, $\rho$, $\eta$, and
$\zeta$, which are the velocity, pressure, areal mass density,
shear (dynamic) viscosity, and second viscosity, respectively, and
$\nu_d$ is the gas friction. The ratio $\eta/\rho$ is the
kinematic viscosity, which has the same dimensions in 2D and 3D
systems. Our flow has a symmetry $\partial /\partial x=0$ and
$v_y=0$. The Navier-Stokes equation is then reduced to $\partial
^2 v_x(y)/\partial y^2-(\nu_d\,\rho/\eta)v_x(y)=0$ and $p=const$.

We will compare our results to a theoretical velocity profile
$v_x^{th}(y)=[(V_1+V_2\,{\rm e}^{2\alpha h}){\rm e}^{\alpha
y}-(V_2+V_1\,{\rm e}^{2\alpha h}){\rm e}^{-\alpha y}]/({\rm
e}^{3\alpha h}-{\rm e}^{-\alpha h})$, where
$\alpha=\sqrt{\nu_d\,\rho/\eta}$ and $2h$ is the distance between
the laser sheets. We used the boundary conditions $v_x(-h)=-V_1$
and $v_x(h)=V_2$; if $V_1=V_2=V$, the theoretical profile
simplifies to $V{\rm sinh}(\alpha y)/{\rm sinh}(\alpha h)$. Here
we have ignored the spatial dependence of $\eta/\rho$, thereby
neglecting the temperature gradient in the shear flow. As
$\nu_d\rightarrow0$, the solution approaches a linear velocity
profile, as for a planar Couette flow. The gas drag $\nu_d\neq0$
gives the steady-state velocity profiles a curvature depending on
viscosity; this allows us to calculate the viscosity from the
profiles, if $\nu_d$ is known.

We fit our experimental velocity profiles in Fig.~\ref {Fig. 2}(c)
to theory using experimental values of $V_1$ and $V_2$ and a
single free parameter $\alpha$. We then calculated the kinematic
viscosity $\eta/\rho=\nu_d\,\alpha^{-2}$ using the known value of
Epstein gas drag $\nu_d=0.87$~s$^{-1}$ for our experimental
conditions \cite {Liu03}. The resulting curves fit our profiles
well.

Our main experimental result is the kinematic viscosity
$\eta/\rho$ of our particle suspension, Fig.~\ref {Fig. 3}(a). Its
value is of order $1~\rm{mm}^2\rm{s}^{-1}$, which is comparable to
$\eta/\rho$ for both a 3D Yukawa system and liquid water \cite
{Murillo01,Salin02&03,Hamaguchi02,Murillo03,Morfill04}. For a
given value of $\kappa$, the parameter we varied was the applied
laser power; increasing this power caused the shear stress to
increase and ${\it \Gamma}_y$ to decrease, so that these two
parameters were not varied independently.

A prominent feature of Figs.~\ref {Fig. 3}(a) and \ref {Fig. 3}(b)
is a broad minimum in the viscosity. The minimum occurs in the
range $70<{\it \Gamma}_y<700$. In 3D Yukawa systems, there is also
a minimum, and the value of ${\it \Gamma}$ for this minimum
depends on $\kappa$ \cite
{Murillo01,Salin02&03,Hamaguchi02,Murillo03}.

We next consider some of the assumptions in our model and validate
them using our experimental results. The weak dependence of
viscosity on ${\it \Gamma}_y$ in the range $70<{\it \Gamma}_y<700$
validates our neglecting the temperature gradient, but only in
this range of ${\it \Gamma}_y$; our data points outside this range
are less reliable, due to this assumption. The Reynolds number of
our shear flow was $R=V_{1}h/(\eta/\rho)=0.7-17$; values this low
typically indicate a laminar flow, validating that assumption in
our model.

We have also assumed that the neutral gas does not flow and
entrain the particles. This is true because the gas-gas collision
mean free path of $15$~mm was of the order of the entire
suspension's size, and the particles filled only a tiny solid
angle. Thus, a gas atom that was struck by one moving particle was
unlikely to strike another.

The good fit in Fig.~\ref {Fig. 2}(c) suggests that the
Navier-Stokes model, which is a continuum model that does not
describe motion of individual molecules, works well even when the
ratio of the shearing region width $2h$ to the Wigner-Seitz radius
$a$ is as small as $17$ to $24$, as it was in our experiment. In
Ref.~\cite {Murillo01}, tests suggested that validity of the
Navier-Stokes model requires that the length scale $2h$ of the
system exceed both the interparticle mean free path $\lambda_{\rm
mfp}$ and $r_{\rm corr}$. Both of these conditions are satisfied
for our experiment. We validate the $2h>\lambda_{\rm mfp}$
condition by observing that $2h/a$ was in the range $17-24$ in our
experiment, and judging that $\lambda_{\rm mfp}\approx a$, based
on the movies of the particle motion \cite {Movie_URL}. We
validate the $2h>r_{\rm corr}$ condition using the pair
correlation results, yielding $2h/r_{\rm corr}=3.0-9.4$ for all
data points except for two with $2h/r_{\rm corr}=2.5-2.9$; for
these, ${\it \Gamma}_y\geq1000$. The latter two data points, which
are for a mostly solid suspension, are therefore less reliable.

However, the Navier-Stokes equation cannot be used inside the
laser footprints, where the shear stress is applied, because this
region is very narrow, less than an interparticle spacing. For
this reason, we did not extend our fits into the footprints. In
general, if the Navier-Stokes equation fails to apply, one cannot
expect the usual relation, that the shear stress is the product of
the dynamic viscosity and shear rate, to apply.

Our experiment suggests a need for a theory or simulation for the
viscosity of a 2D Yukawa system. The only previous work, to our
knowledge, is for a 3D Yukawa system \cite
{Murillo01,Salin02&03,Hamaguchi02,Murillo03}.

After we submitted this paper we learned of the experiment of
Ref.~\cite {Fortov04}, where the shear viscosity of a
multi-layered dusty plasma was measured using a shear flow induced
by a single laser beam.

We thank B.~Liu, J.~Marshall, R.~Merlino, and F.~Skiff for
valuable discussions. This work was supported by NASA and DOE.

\begin{figure}[p]
\caption {Particle suspension at three levels of shear stress
applied by a pair of counter-propagating laser sheets in a planar
Couette configuration. Panels (a), (c), and (e) are top views with
shear stress that is zero, moderate, and high, respectively. The
snapshot of the undisturbed suspension (a) shows a highly ordered
triangular lattice. At the onset of plastic flow (c), particle
trajectories reveal localized hopping. In a fully developed shear
flow (e), the trajectories show particles moving significantly
everywhere. Profiles of the inverse particle temperature
$T_x^{-1}$ and $T_y^{-1}$ are shown in the insets. The static
structure factor $S(\bf{k})$ was computed as the Fourier
transform, with a Hanning window, of the raw bitmap images as in
(a). The peaks in $S(\bf{k})$ are distinctive for an undisturbed
lattice (b); they are broadened and then diffuse at progressively
higher levels of shear (d),(f). The particle suspension's
parameters here and in Fig.~\ref {Fig. 2} correspond to
Experiment~1 in Table~1. We calculated $\eta$, $<T_y>$, and ${\it
\Gamma}_y$ using data in the range $-h<y<h$.} \label {Fig. 1}
\end{figure}

\begin{figure}[p]
\caption {Particle velocity data. Map (a), averaged over $64$
frames at the onset of plastic flow, shows particle movement
localized near the laser footprints. Map (b), averaged over $256$
frames at the highest laser power, shows a fully developed shear
flow that has no variation with $x$. Experimental profiles (data
points) (c) are with fits to the Navier-Stokes model (curves), as
used to calculate the kinematic viscosity of the particle
suspension.} \label {Fig. 2}
\end{figure}

\begin{figure}[p]
\caption {Results for kinematic viscosity $\eta/\rho$, as
functions of the shear stress (a) and coupling parameter ${\it
\Gamma}_y$ (b). The latter two parameters were not varied
independently, but were related as shown in (c). There is a broad
minimum in viscosity in the range $70<{\it \Gamma}_y<700$. In (b),
to allow comparison to theory, we normalized the viscosity as
$\eta^*=\eta/(\rho\omega_{pd}\,a^2)$, where
$\omega_{pd}\equiv[Q^2/2\pi\epsilon_0ma^3]^{1/2}$ for two
dimensions \cite {Kalman04}.} \label {Fig. 3}
\end{figure}

\begin{table}
\caption{Parameters of the particle suspension measured without
any shear flow, for three experiments.}
\begin{tabular}{lccc}
&1&2&3\\ \tableline Wigner-Seitz radius $a$ (mm)
&0.402&0.449&0.580\\ $\kappa$ &0.36&0.42&0.53\\ $Q/e$
&-11~940&-13~840&-16~360\\
$\omega_{pd}\equiv[Q^2/2\pi\epsilon_0ma^3]^{1/2}$
(s$^{-1}$)&49.2&48.2&38.8\\
\end{tabular}
\end{table}

\end{document}